\documentstyle[graphicx,aps,amstex,epsf]{revtex}
\begin{document} 
\textheight 23 cm \topmargin -1 cm 
\textwidth 15 cm 
\setlength{\baselineskip}{0.333333in} 
\draft 
\preprint{ }
\title{Skyrmion Physics in Bose-Einstein Ferromagnets}
\author{U.  Al Khawaja and H.  T.  C.  Stoof} \address{Institute
for Theoretical Physics, University of Utrecht, Princetonplein 5,3584 CC
Utrecht, The Netherlands} 
\date{\today} 
\maketitle 
\begin{abstract}
We show that a ferromagnetic Bose-Einstein condensate has not only line-like 
vortex excitations, but in general also allows for point-like topological
excitations, i.e., skyrmions. We discuss the thermodynamic stability and the 
dynamic properties of these skyrmions for both spin-1/2 and ferromagnetic 
spin-1 Bose gases. 
\end{abstract}


\section{Introduction}

An understanding of quantum magnetism is important for a
large number of phenomena in physics. Three well-known examples are
high-temperature superconductivity, quantum phase transitions and the quantum
Hall effect. Moreover, it appears that magnetic properties are also very
important in another area, namely Bose-Einstein condensation in trapped atomic
gases. This has come about because of two independent experimental
developments. The first development is the realization of an optical trap for
atoms, whose operation no longer requires the gas to be doubly spin-polarized
\cite{MIT1,MIT2}. The second development is the creation of a two-component 
Bose-Einstein condensate \cite{JILA1}, which by means of rf-fields can be manipulated so as to
make the two components essentially equivalent \cite{JILA2}. As a result the
behavior of both spin-1 and spin-1/2 Bose gases can now be experimentally
explored in detail. Indeed, at present already such diverse phenomena as domain 
walls \cite{domains}, 
macroscopic quantum tunneling \cite{tunneling}, Rabi oscillations 
and vortices \cite{JILA21} have been observed.

Theoretically, the ground-state structure of these gases has recently been
worked out by a number of authors \cite{jason1,OM,nick,jason2} and also the 
first studies of the line-like vortex excitations have appeared 
\cite{jason1,Yip,TK}.
However, an immediate question that comes to mind is whether the spin degrees of
freedom allow for other topological excitations that do not have an analogy in
the case of a single component or scalar Bose condensate. It is one of the aims of this
paper to show that the answer to this question is in general affirmative. In
particular, we show that a ferromagnetic Bose-Einstein condensate has so-called
skyrmion excitations, which are nonsingular but nevertheless topologically 
nontrivial point-like spin textures. Roughly speaking, the skyrmion is an 
excitation that can be created out of the ground state, in which all the spins 
are aligned, by reversing the average spin in a finite region of space. 
Skyrmions are also known from nuclear physics \cite{Skyrme} and the quantum Hall 
effect \cite{sondhi}, but to observe them in an atomic gas would be exciting 
since in that case a completely microscopic understanding of their behavior is 
possible. In nuclear physics and the quantum Hall effect this is not true 
because of the nonperturbative nature of QCD and the presence of impurities that 
obstruct the center-of-mass motion of the skyrmions, respectively.  Having 
proven their existence, we then turn to the investigation of the precise texture, 
the energetic 
stability, and finally the dynamical behavior of skyrmions.
Some of the results of this analysis we have already reported in previous 
communications \cite{us1}, but here we try to give a much more complete and 
detailed picture of the skyrmion physics in spinor Bose-Einstein condensates.

With this in mind, we would like to mention that we recently have also 
considered 't Hooft-Polyakov monopoles \cite{mono} in an antiferromagnetic 
Bose-Einstein condensate \cite{us2}. These topological excitations are in fact 
singular skyrmions, but have quite different properties than the nonsingular 
skyrmions which are the object of study in this paper. In particular, due to the 
singular nature of the spin texture of the 't Hooft-Polyakov monopole, the 
condensate density vanishes in the core and the monopole turns out to be 
thermodynamically stable. This is completely analogous to the case of a vortex 
in a scalar Bose-Einstein condensate. Both these features are not shared by the 
nonsingular skyrmions, which complicates the analysis considerably. The most 
important problem in this respect is that for a nonsingular skyrmion the 
topology allows for a spin texture with an arbitrary intrinsic size. As a result 
the stability of the skyrmion is now determined by energetic arguments and not by 
topological arguments as in the case of the singular 't Hooft-Polyakov monopole.

The paper is organized as follows. In Sec. \ref{defining}, we use the symmetry 
properties of the order parameter of a spinor Bose-Einstein condensate 
to show that quite generally skyrmion excitations indeed exist in such a 
condensate. We then turn our attention to the ferromagnetic case and discuss 
some general properties of these 
topological excitations, especially the spin texture, the superfluid velocity 
profile, and also the density profile. 
In Sec. \ref{stability} we investigate the energetic 
stability of the skyrmion and show that from a thermodynamic point of view they 
are always unstable and tend to shrink to microscopic sizes. As a result we then 
consider the dynamical stability of the skyrmion and in particular determine
the rate at which they collapse. We find that under certain conditions this rate is actually much 
smaller 
than the decay rate of the condensate itself due to various inelastic processes. Therefore, the 
skyrmion 
can for all practical purposes be considered as a (meta)stable excitation and we are justified to look 
in 
Sec. \ref{dynamicproperties} also at other important dynamic properties 
of the skyrmion such as the ``spin'' and the center-of-mass motion. 
Finally, we end in Sec. \ref{conclusion} by a summary and some conclusions.

\section{Skyrmions as Topological Excitations}
\label{defining}
In this section we discuss in detail the main static features of skyrmions. 
First of all, we show from the symmetry of the order-parameter space of the 
spinor condensate, that from a topological 
point of view point-like skyrmion excitations can 
indeed exist in both spin-1/2 and spin-1 Bose-Einstein 
condensates. Focussing then on the ferromagnetic case, we introduce a 
convenient parameterization of the skyrmion texture, which allows us to 
incorporate most easily the nontrivial winding number associated with the 
skyrmion. Next, we write down the energy functional 
for a ferromagnetic spinor condensate and, by substituting the above mentioned 
parametrization, derive the corresponding Euler-Lagrange equations for the 
density profile and the spin texture of the skyrmion. 
Finally, to simplify the actual calculation of the skyrmion density profile and 
the spin texture, we propose a variational approach that automatically takes into account 
the desired overall features of the skyrmion texture. 

\subsection{Topological Considerations}
\label{topological}
To find all possible
topological excitations of a spinor condensate, we need to know
the full symmetry of the macroscopic wave function $\Psi({\bf r})
\equiv \sqrt{n({\bf r})} \zeta({\bf r})$, where $n({\bf r})$ is
the total density of the gas, $\zeta({\bf r})$ is a normalized
spinor that determines the average local spin by means of $\langle
{\bf S} \rangle({\bf r})
    = \zeta^*_{a}({\bf r}) {\bf S}_{ab} \zeta_{b}({\bf r})$,
and ${\bf S}$ are the usual spin matrices obeying the commutation relations
$[S_{\alpha},S_{\beta}]
                    = i \epsilon_{\alpha\beta\gamma} S_{\gamma}$.
Note that here, and in the following, summation over repeated
indices is always implicitly implied. From the work of Ho
\cite{jason1} we know that in the case of spin-1 bosons we have to
consider two possibilities, since the effective interaction
between two spins can be either antiferromagnetic or
ferromagnetic. In the antiferromagnetic case the ground-state
energy is minimized for $\langle {\bf S} \rangle({\bf r}) = {\bf
0}$, which implies that the parameter space for the spinor
$\zeta({\bf r})$ is only $U(1)\times S^2$ because we are free to
choose both its overall phase and the spin quantization axis. In
the ferromagnetic case the energy is minimized for $|\langle {\bf
S} \rangle({\bf r})| = 1$ and the parameter space corresponds to
the full rotation group $SO(3)$. Using the same arguments, we find
that for spin-1/2 bosons the order-parameter space of the ground
state is always equivalent to $SU(2)$ \cite{scat}.

What do these results tell us about the possible topological
excitations \cite{raja,mermin}? For line-like defects or vortices,
we can assume $\zeta({\bf r})$ to be independent of one direction
and the spinor represents a mapping from a two-dimensional plane
into the order-parameter space. If the vortex is singular this
will be visible on the boundary of the two-dimensional plane and
we need to investigate the properties of a continuous mapping from
a circle $S^1$ into the order-parameter space $G$, i.e., of the
first homotopy group $\pi_1(G)$. Since $\pi_1(SU(2)) =
\pi_1(SO(3)) = Z_2$ and $\pi_1(U(1)\times S^2) = Z$, we conclude
that a spin-1/2 and a ferromagnetic spin-1 condensate can have
only vortices with a winding number equal to 1, whereas an
antiferromagnetic spin-1 condensate can have vortices with winding
numbers that are an arbitrary integer. Physically, this means that
by traversing the boundary of the plane, the spinor can wind
around the order-parameter at most once or an arbitrary number of
times, respectively. This conclusion is identical to the one
obtained previously by Ho \cite{jason1}.

If the vortex is nonsingular, however, the spinor $\zeta({\bf r})$
will be identical everywhere on the boundary of the
two-dimensional plane and it effectively represents a mapping from
the surface of a three-dimensional sphere $S^2$ into the order
parameter space. We then need to consider the second homotopy
group $\pi_2(G)$. For this we have that $\pi_2(SU(2)) =
\pi_2(SO(3)) = 0$ and $\pi_2(U(1) \times S^2) = Z$. Hence
nonsingular or coreless vortices are only possible for a spin-1
condensate with antiferromagnetic interactions. It therefore
appears that the nonsingular spin texture discussed in Ref.
\cite{jason1}, is topologically unstable and can be continuously
deformed into the ground state by ``local surgery'' \cite{mermin}.

We are now in a position to discuss point-like defects. Since the boundary of a
three-dimensional gas is also the surface of a three-dimensional sphere,
singular point-like defects are also determined by the second homotopy group
$\pi_2(G)$. Such topological excitations thus only exist in the case of a
spin-1 Bose gas with antiferromagnetic interactions. We call these excitations 
't Hooft and Polyakov monopoles \cite{mono}, although it
would be justifiable to call them singular skyrmions. For nonsingular
point-like defects the spinor $\zeta({\bf r})$ will again be identical on the
boundary of the three-dimensional gas. As a result, the configurations space is
compactified to the surface of a four-dimensional sphere $S^3$ and we need to
determine the third homotopy group $\pi_3(G)$. For this we find $\pi_3(SU(2)) =
\pi_3(SO(3)) = \pi_3(U(1) \times S^2) = Z$. Hence nonsingular skyrmion
excitations exist in all three cases.

\subsection{Skyrmion Texture}
\label{texture}
We consider from now on only the case of a homogeneous and ferromagnetic spinor 
condensate. In the ground state all spins are aligned 
along the direction of a uniform and sufficiently weak magnetic field, which we 
take to be along the $z$-axis. The uniform magnetic field is needed only to direct the 
spins in the ground state, but it 
should not provide a substantial energy barrier 
for spin flips. The fact that we consider a homogeneous gas and not a 
confined 
one is only for simplicity and turns out not to be crucial for the practical applicability 
of our work. This 
is so because the, for our purposes relevant length scale over which the skyrmion 
spin deformations take place is always of the 
order of 
the correlation length, which under typical experimental conditions is much less 
than the length scale for density gradients due to the confining potential.   
The skyrmion excitation is a space-dependent spin deformation of the ground 
state and can thus be represented by a position-dependent spinor 
$\zeta({\bf r})$.
A convenient way of introducing the position-dependence in the spinor 
is to write it in terms of a position-dependent rotation that acts on the 
constant 
spinor $\zeta^{\rm Z}$ associated with the ferromagnetic groundstate. In this manner 
we have
\begin{equation}
\zeta({\bf r})=\exp{\left\{-{i\over S}\mbox{\boldmath $\Omega$}({\bf r})\cdot 
{{\bf S}}\right\}}\zeta^{\rm Z}
\label{spinor}.
\end{equation}
Here, the constant spinor $\zeta^{\rm Z}$ minimizes the Zeeman energy and is 
given by
\begin{equation}
\zeta^{\rm Z}=
\left\{
\begin{array}{cr}
\left(\begin{array}{c}
1\\0
\end{array}\right),&S={1\over2}
\\
\\
\left(\begin{array}{c}
1\\0\\0
\end{array}\right),&S=1
\end{array}\right.
\label{zetaz} 
\end{equation}
in the usual basis that diagonalizes $S_z$. Furthermore, $\mbox{\boldmath $\Omega$}({\bf r})$ 
is a real vector function of $\bf r$. 
It parametrizes the ferromagnetic order-parameter space, which due to our incorporation of the factor 
$1/S$ 
in Eq. (\ref{spinor}) is always a sphere 
of radius $\pi$. We point out that here and throughout the following $S$  can only take the values 1/2 
or 1.  The significance of Eq.  (\ref{spinor}) is that at a point $\bf r$ 
the spinor $\zeta^{\rm Z}$ is rotated by an angle 
that equals $|\mbox{\boldmath $\Omega$}({\bf r})|/S$ around the  direction of  
$\mbox{\boldmath $\Omega$}({\bf r})$. There is no restriction on the generality 
of spin textures produced by this means; it is merely a convenient 
parameterization of the order-parameter space in terms of 
$\mbox{\boldmath $\Omega$}({\bf r})$. 
Since we are mostly interested in the equilibrium properties, we assume here for 
simplicity the maximally symmetric shape of the skyrmion 
which is expected to have the smallest possible gradients. This means that we take 
\begin{equation}
{\mbox{\boldmath $\Omega$}}({\bf r})=\omega(r){\bf r}/r\equiv\omega(r){\hat{\bf r}}
\label{simplicity},
\end{equation}
where the function $\omega(r)$ should obey the boundary conditions 
$\omega(0)=2\pi$ and $\lim_{r\rightarrow\infty}\omega(r)=0$ \cite{sh}. 
Thus, at these boundaries the rotation operator in Eq. (\ref{spinor})  
becomes the identity and gives rise to $\zeta({\bf r}={\bf 0})=\zeta({\bf 
r}\rightarrow\infty)=\zeta^{\rm Z}$.
Furthermore, $\omega(r)$ should as a function of radius decrease monotonocally from $2\pi$ to 0, 
since this will correspond to the smallest gradient energy for the spin deformations. 
With this ansatz for ${\mbox{\boldmath $\Omega$}}({\bf r})$ and its boundary 
conditions, 
we see that by traversing the whole
configuration space, we exactly cover the order-parameter space
twice, which is required to avoid a singular behavior of the
spinor at ${\bf r} = {\bf 0}$. 
Indeed, the boundary condition $\omega(0)=\pi$, which in first 
instance appears to be the right 
one as it leads to a spin texture which covers the order-parameter space only once, is physically 
unacceptable because it results in a singular behaviour of the spinor $\zeta({\bf r})$ and therefore in 
divergencies in the gradient energy. The reason for this singular behaviour is that, given the 
maximally 
symmetric shape of the skyrmion, the spinor $\zeta({\bf r})$ is always equal to $\zeta^{\rm Z}$ on the 
$z$-axis.
Note that apart from topological reasons, the condition $\lim_{r\rightarrow\infty}\omega(r)=0$ also 
ensures 
that the spin deformations associated with the skyrmion have a finite 
range, so that only a finite amount of energy is required to excite the skyrmion 
from the ground state. Thus, both on the $z$-axis and far away from the center of the skyrmion the 
spins are 
directed as in the ground state. From these general features we can now draw a qualitative 
picture of the skyrmion texture. Suppose we approach the origin along a line through the origin, 
starting 
from a distance much larger than the range of 
the spin deformations. Then how does the average local spin 
behave as a function of distance? 
According to the above description of the rotation operator depicted in Fig.  \ref{schematic}, 
the spin initially points up along the 
$z$-axis, but by approaching the 
origin it will start to rotate around the radial direction until we approach the origin and the spin 
points 
up again. It is essential that the spins complete an integer number of cycles equal to $1/S$ in order 
to have a nonzero topological winding number given by 
\begin{equation}
N^{\rm sk} = \frac{3}{8\pi^4} \int d{\mbox{\boldmath $\Omega$}}
 = \frac{1}{16\pi^4} \int d{\bf r} \epsilon^{ijk} \epsilon_{\alpha\beta\gamma}
      \partial_i \Omega^{\alpha} \partial_j \Omega^{\beta}
      \partial_k \Omega^{\gamma}
\label{winding},
\end{equation}  
which in this case is equal to 1.

Summarizing, we have obtained the desired position dependence of $\zeta({\bf r})$ in terms of a single 
function $\omega(r)$ that describes how the average local spin vector 
is 
tilted from its orientation in the ground state. This function and the local 
density 
$n({\bf r})$ represent the two degrees of freedom that describe the skyrmion and our next 
task is to 
determine their precise spacial dependence.

\subsection{Energy Functional}
\label{energy functional}
For a ferromagnetic spinor condensate, Ho \cite{jason1} has shown that, 
within the mean-field approach, the energy 
functional in the absence of a magnetic field is given by
\begin{eqnarray}
E[n,\zeta]&\equiv&\int d{\bf r}\left[
{\hbar^2\over 2m}\left(\mbox{\boldmath $\nabla$}{\sqrt{n({\bf r})}}\right)^2
-\mu n({\bf r})\right.\nonumber\\
&+&\left.{\hbar^2\over 2m}n({\bf r})|\mbox{\boldmath $\nabla$} \zeta ({\bf 
r})|^2
+{1\over2}T^{2B}n^2({\bf r})\right]
\label{functional},
\end{eqnarray}                                                                                     
where $m$ is the mass of the atoms, $\mu$ is their chemical potential and  
$T^{2B}=4\pi a\hbar^2/m$ 
is the appropriate coupling constant that represents the strength of 
the interatomic interactions in terms of the positive scattering 
length $a$. 
The term $\hbar^2n({\bf r})|\mbox{\boldmath $\nabla$} \zeta ({\bf r})|^2/2m$  
represents the energy density associated with the gradients in the spin texture. It is this term that 
enforces the boundary condition $\omega(0)=2\pi$, as discussed previously.

For sufficiently large distances the gradient terms vanish and we infer 
from Eq. (\ref{functional}) that $\mu=n_{\infty}T^{2B}$, 
where $n_{\infty}=n({\bf r}\rightarrow\infty)$. 
Using this, Eq. (\ref{functional}) can be put in a dimensionless form by 
scaling the lengths to the correlation length $\xi=1/\sqrt{8\pi a n_{\infty}}$, the density to 
$n_{\infty}$, and the total energy to $\hbar^2n_{\infty}\xi/2m$. In this manner Eq. (\ref{functional}) 
becomes
\begin{equation}
\varepsilon [f,\zeta]\equiv\int d\mbox{\boldmath $\rho$}\left[
 \left(\mbox{\boldmath $\nabla$}_{\mbox{\boldmath $\rho$}}
 {\sqrt{f(\mbox{\boldmath $\rho$})}}\right)^2
+f(\mbox{\boldmath $\rho$})|\mbox{\boldmath $\nabla$}_{\mbox{\boldmath $\rho$}} 
\zeta (\mbox{\boldmath $\rho$})|^2
+{1\over2} f^{2}(\mbox{\boldmath $\rho$})-f(\mbox{\boldmath $\rho$})\right]
\label{fs},
\end{equation}
with $\mbox{\boldmath $\rho$}={\bf r}/\xi$ and $n({\bf r})=n_{\infty}f({\mbox{\boldmath $\rho$}})$.
The gradient term $|\mbox{\boldmath $\nabla$}_{\mbox{\boldmath $\rho$}}\zeta(\mbox{\boldmath 
$\rho$})|^{2}$ is calculated 
in Appendix 
\ref{a} explicitly 
as a function of $\omega(\rho)$. 
This is achieved by inserting in the usual basis the spin-1 and spin-1/2 matrices 
in the rotation operator, performing a power expansion and then 
using some properties of the powers of the spin matrices to sum the resulting infinite 
series 
into the following compact formula
\begin{equation}
|\mbox{\boldmath $\nabla$}_{\mbox{\boldmath $\rho$}} \zeta(\mbox{\boldmath $\rho$})|^2
=\left\{
\begin{array}{ll}
2\left({\sin{(\omega(\rho))}\over \rho}\right)^2
+\left({d \omega(\rho)\over d\rho}\right)^2, & S=1/2\\
(5-\cos{(2\theta)})
\left({\sin{(\omega(\rho)/2)}\over \rho}\right)^2
+{1\over4}(3+\cos{(2\theta)})\left({d \omega(\rho)\over d\rho}\right)^2, & S=1
\end{array}\right.
\label{gradzeta2},
\end{equation}  
where $\theta$ is the azimuthal angle between $\mbox{\boldmath $\rho$}$ and the $z$-axis. 

Finally, the density profile $f(\mbox{\boldmath $\rho$})$ and the function $\omega(\rho)$ can be 
determined from the two coupled differential equations resulting from 
minimizing $\varepsilon[f, \zeta]$ with respect to $f(\mbox{\boldmath $\rho$})$ 
and 
$\omega(\rho)$, namely
\begin{equation}
-{\mbox{\boldmath $\nabla$}_{\mbox{\boldmath $\rho$}}^{2}
\sqrt{f(\mbox{\boldmath $\rho$})}\over\sqrt{f(\mbox{\boldmath $\rho$})}}
+|{\bf \mbox{\boldmath $\nabla$}_{\mbox{\boldmath $\rho$}}}
\zeta(\mbox{\boldmath $\rho$})|^{2}
+f(\mbox{\boldmath $\rho$})-1=0,
\label{gp}
\end{equation} 
and
\begin{equation}
{1\over S}
\langle f({\mbox{\boldmath $\rho$}})\rangle_1\sin{\left({\omega(\rho)\over S}\right)}
-2S{d\over d\rho}
\left[
\rho^{2}\langle f(\mbox{\boldmath $\rho$})\rangle_2
\left({d\omega(\rho)\over d\rho}\right)^{2}
\right]=0
\label{omegaequation},
\end{equation}
where
\begin{equation}
\langle f({\mbox{\boldmath $\rho$}})\rangle_1
=\left\{
\begin{array}{lr}
{1\over4\pi}\int d\hat{\mbox{\boldmath $\rho$}} 
f({\mbox{\boldmath $\rho$}}),&S=1/2 \\\\
{1\over4\pi}\int d\hat{\mbox{\boldmath $\rho$}} 
(5-\cos{(2\theta)})f({\mbox{\boldmath $\rho$}}),&S=1 
\end{array}\right.
\end{equation}
and
\begin{equation}
\langle f({\mbox{\boldmath $\rho$}})\rangle_2
=\left\{
\begin{array}{lr}
{1\over4\pi}\int d\hat{\mbox{\boldmath $\rho$}} 
f({\mbox{\boldmath $\rho$}}),&S=1/2 \\\\
{1\over4\pi}\int d\hat{\mbox{\boldmath $\rho$}} 
{1\over4}(3+\cos{(2\theta)})f({\mbox{\boldmath $\rho$}}),&S=1 
\end{array}
\right.~.
\end{equation}
In the spin-1 case, Eqs. (\ref{gradzeta2}) and (\ref{gp}) show that the density profile is in principle 
anisotropic. However, the anisotropy turns out to be rather small as we show explicitly in Sec. 
\ref{stability}.

\subsection{Ansatz for $\omega(r)$}
\label{ansatz_section}

\label{profile}
In principle, to calculate the exact skyrmion texture and density profile, we  
should now solve the two coupled nonlinear differential equations that result 
from minimizing the energy functional with respect to $f(\mbox{\boldmath $\rho$})$ and 
$\zeta(\mbox{\boldmath $\rho$})$, namely Eqs. (\ref{gp}) and (\ref{omegaequation}). 
As an alternative, we employ here a rather simpler, though less rigorous, approach. 
We simplify the calculations by introducing an ansatz for $\omega(\rho)$ that takes 
explicitly into account the physical boundary conditions discussed in the previous section. 
Our ansatz is
\begin{equation}
\omega(\rho)=4\cot^{-1}{(\rho/\lambda)^{2}}
\label{omegaansatz},
\end{equation}
where $\lambda$ is a parameter that determines the radius at which $\omega(\rho)$  
crosses over from $2\pi$ to 0 and physically corresponds to the size of the skyrmion.
We see that this ansatz automatically satisfies the boundary conditions 
$\omega(0)=2\pi$ and $\lim_{\rho\rightarrow \infty}\omega(\rho)=0$, as 
required. 
It should be noted here that the detailed functional behavior of $\omega(\rho)$ 
will turn out not to be crucial for our results on the skyrmion stability as long as it 
satisfies 
the prescribed boundary conditions and falls of monotonically. 
To substantiate this remark, we have performed also calculations using a number of 
different 
forms for 
$\omega(\rho)$ that satisfy the desired boundary conditions such as 
$\omega(\rho)=2\pi/(1+(\rho/\lambda)^2)$. We 
found, as expected, that the skyrmion energy only differs slightly from one function to the other and, 
in particular, that the energy is minimized for the function 
given in 
Eq. (\ref{omegaansatz}). Furtunately, this ansatz is also simpler to 
handle analytically.

\section{Static and dynamical stability}
\label{stability}

The most important question about the skyrmion excitation 
is whether it is energetically stable or not. In other words, how does the 
energy of the 
skyrmion 
depends on its size? As we show next, the answer to this question is that 
the skyrmion always tends to shrink to microscopic sizes to minimize its energy. 
Although this is an unfortunate result, it does not need to rule out an experimental 
observation of the skyrmion excitation as long as the typical time scale 
for this collapse is sufficiently long. 
Therefore, we also consider this problem after we have discussed the thermodynamical 
stability of the skyrmion.

For sufficiently large skyrmions the gradients of the spinor $\zeta({\bf r})$ 
are small and density fluctuations are therefore also small. The 
energy of the 
skyrmion 
can then be approximated by 
$({\hbar^2n_{\infty}/2m})\int d{\bf r}|\mbox{\boldmath $\nabla$}\zeta({\bf r})|^2$.  
If the size of the skyrmion is of order of $\lambda$, this energy scales as 
$\lambda$. 
This 
indicates that the skyrmion tends to shrink in order to minimize its energy. 
However, for 
smaller 
sizes the density fluctuations and their gradients start to grow and 
this simple 
argument 
no longer applies. For large skyrmions Eq. (\ref{gp}) shows that the density 
fluctuations scale 
as 
$f(\rho)-1\approx|\mbox{\boldmath $\nabla$}_{\mbox{\boldmath $\rho$}}
\zeta(\mbox{\boldmath $\rho$})|^2\propto1/\lambda^2$ 
and their energy contribution thus behaves as $1/\lambda$. Approaching 
smaller values of $\lambda$, the kinetic energy term in Eq. (\ref{gp}) increases and the 
density 
fluctuations 
will scale with a power of $\lambda$ that is greater than $-2$, 
because otherwise the density would become negative at some point. 
As a result the energy will 
scale with a power of $\lambda$ that is different from 1. 
Therefore, there is in principle a chance for stability if 
the energy associated with the 
density fluctuations is at some point increasing when $\lambda$ becomes smaller. 
To investigate 
this possibility we need to calculate the energy as a function of 
$\lambda$ exactly 
for all 
values of $\lambda$.
If the energy function possesses a (local) global minimum 
for a finite $\lambda$, then the skyrmion is energetically 
(meta)stable. 
We have indeed calculated this energy curve, 
the details can be found in Sec. \ref{baby} below, and it turns out that 
the skyrmion energy actually increases monotonically with $\lambda$. 
This means that a skyrmion of any finite size is energetically unstable and 
will tend to shrink to zero size. 
Of course, this condition holds within the Gross-Pitaevskii 
theory which describes only the long-wavelength physics. 
Corrections to this theory will lead to a finite, but 
microscopically small size of the skyrmion. 
However, as mentioned previously, for observing the skyrmion experimentally 
it is important to know the time scale for its collapse 
since it could be larger than the lifetime of the condensate itself 
due to various inelastic processes, 
such as two- and three-body collisions and, 
in an optical trap, photon absorption. 

From our analysis of the skyrmion shrinking rate, it turns out that there 
are two size regimes with different shrinking mechanisms. 
The first is for skyrmions with a size $\lambda$ that is much larger than the 
correlation length of the gas. 
We denote such large skyrmions as ``adult'' skyrmions. 
The second regime is for much smaller skyrmions with sizes of the order of 
the correlation length or less. We denote these small skyrmions as ``baby'' 
skyrmions. 
In the next two subsections we calculate the skyrmion energy 
and estimate its shrinking rate in these two regimes.

\subsection{Adult Skyrmions}
\label{adult}
As mentioned earlier, adult skyrmions are large skyrmions with a size $\lambda$ that 
satisfies 
$\lambda\gg\xi=1/\sqrt{8\pi a n_{\infty}}$, 
where $\xi$ is the correlation length of a homogeneous 
gas of average density $n_{\infty}$ 
and $s$-wave interatomic scattering length $a$.
In this section we express the lagrangian of the skyrmion in terms of a 
time-dependent skyrmion size $\lambda(t)$. 
An equation of motion for $\lambda(t)$ is then derived. 
It should be noted here that comparison between our calculation in this section 
and experiment is only meaningful 
for skyrmion sizes much less than the condensate size since this 
calculation
is for a homogeneous system and does not take into account the effect of the 
trap. 
The two conditions thus imply that the spinor condensate is deep in the Thomas-Fermi regime.

For adult skyrmions the gradients in the spinor $\zeta({\bf r})$ are small and therefore the
fluctuations in the density $\delta n({\bf r}) = n({\bf r}) -
 n_{\infty}$ are also small compared to
the average density $ n_{\infty}$. The energy of
the skyrmion can thus be determined by an harmonic approximation to the
Gross-Pitaevskii energy functional given in Eq. (\ref{functional}). 
To second order in $\delta n({\bf r})$
it is given by
\begin{eqnarray}
E[n,\zeta] = \frac{1}{2} \int d{\bf r} \int d{\bf r}'~
     \delta n({\bf r}) \chi^{-1}({\bf r}-{\bf r}')
                        \delta n({\bf r}')         \hspace*{0.3in} \nonumber \\
+ \int d{\bf r}~ n({\bf r})
     \left( \frac{\hbar^2}{2m} |\mbox{\boldmath $\nabla$} \zeta({\bf r})|^2
         - \gamma {\bf B} \cdot \langle {\bf S} \rangle({\bf r})
         \right)~,
\label{eadult}
\end{eqnarray}
where $\chi({\bf r}-{\bf r}')$ is the static density-density
response function, which is defined by
\begin{equation}
\frac{\hbar^2}{4m n_{\infty}}
  \left( -\nabla^2
    + 16\pi a n_{\infty}\right) \chi({\bf r}-{\bf r}')
= \delta({\bf r}-{\bf r}')
\label{response}
\end{equation}
but explicitly reads
\begin{equation}
\chi({\bf r}-{\bf r}')
={m n_{\infty}\over\pi\xi\hbar^2}{\exp{\left(-\sqrt{2}|{\bf r}-{\bf 
r}'|/\xi\right)}\over|{\bf 
r}-{\bf r}'|/\xi}
\label{chie},
\end{equation}
${\bf B}$ is either 
a fictitious, caused by resonant rf-fields, or a real
homogeneous magnetic field, and $\gamma$ is the corresponding magnetic
moment of the atoms in the trap. 
Considering first the ideal case of zero magnetic field
${\bf B}$ and solving for the density fluctuations induced by the spin texture, 
the lagrangian takes the form $L[\zeta]=T[\zeta]-E[\zeta]$, 
where
\begin{eqnarray}
E[\zeta] = \int d{\bf r}~  n_{\infty}
      \frac{\hbar^2}{2m} |\mbox{\boldmath $\nabla$} \zeta({\bf r},t)|^2
             \hspace*{0.8in}  \nonumber \\
- \frac{\hbar^4}{8m^2} \int d{\bf r} \int d{\bf r}'~
      |\mbox{\boldmath $\nabla$} \zeta({\bf r},t)|^2 \chi({\bf r}-{\bf r}')
      |\mbox{\boldmath $\nabla$} \zeta({\bf r}',t)|^2~,
\label{egrad}
\end{eqnarray}
and
\begin{eqnarray}
T[\zeta] &=& \int d{\bf r}~  n_{\infty}
      {\zeta^{\dagger}}({\bf r},t)i\hbar\frac{\partial}{\partial t} \zeta({\bf 
r},t)
             \nonumber \\
&+&{1\over2}\int d{\bf r} \int d{\bf r}'~
       {\zeta^{\dagger}}({\bf r},t)i\hbar\frac{\partial}{\partial t}\zeta({\bf 
r},t)
         \chi({\bf r}-{\bf r}')
        {\zeta^{\dagger}}({\bf r}',t)i\hbar\frac{\partial}{\partial t}\zeta({\bf 
r}',t)
\label{t},
\end{eqnarray}
as a result of the fact that in Gross-Pitaevskii theory, the action for the ferromagnetic condensate 
contains the time-derivative term $\int d t\int d{\bf r}\;n({\bf r},t)\;\zeta^{\dagger}({\bf 
r},t)\;i\hbar\;\partial\zeta({\bf r},t)/\partial t$.  
The equation of motion for $\lambda(t)$ can be 
derived from the above lagrangian by substituting our ansatz for 
$\zeta({\bf r},t)$ given in Eqs. (\ref{spinor}), (\ref{simplicity}), and 
(\ref{omegaansatz}). 
In the present limit, $\lambda\gg\xi$ and the second term in the energy $E[\zeta]$ 
is much smaller than the first one. Therefore, we do not take 
it into 
account. 
Using the explicit expressions for $|\mbox{\boldmath $\nabla$}\zeta({\bf 
r})|^{2}$ 
from Eq. (\ref{gradzeta2}), 
the energy is calculated to be
\begin{equation}
E[\lambda]=C{\hbar^{2}\over2m}n_{\infty}\lambda 
\label{eadult2},
\end{equation}    
where $C=(29-20S)\sqrt{2}\pi^{2}$. 
For the time-dependent contribution $T[\zeta]$ we calculate first 
$\zeta^{\dagger}({\bf r}, t){\partial\zeta({\bf r}, t)/\partial t}$ which turns 
out to be 
equal to $-i{\dot\lambda}({\partial\omega/\partial\lambda})\cos{\theta}$ for both 
$S=1/2$ 
and $S=1$, where the dot denotes a derivative with respect to time. 
Since ${\partial\omega/\partial\lambda}$ is an even function in $r$, the 
first term of $T[\zeta]$ vanishes. Thus $T[\lambda]$ turns out to 
be
\begin{equation}
T[\lambda]={\lambda\over\sqrt{2}a}m{\dot\lambda}^{2}
\label{tadult}.
\end{equation}
So combining Eqs. (\ref{eadult2}) and (\ref{tadult}), we see that the action for the dynamical 
variable $\lambda(t)$ is equivalent to that of a particle with a position-dependent effective 
mass $m^{*}={\sqrt{2}m\lambda/a}$ in the linear potential 
$V(\lambda)=C\hbar^2n_{\infty}\lambda/2m$. 
Scaling again $\lambda$ to the correlation length $\xi$, the number 
${\sqrt{2}\lambda/a}$ 
takes the form $\lambda/\sqrt{4\pi n_{\infty}a^{3}}$. Experimentally, 
the dimensionless parameter $n_{\infty}a^{3}$ is typically of order $10^{-5}$. 
This leads to 
an effective mass $m^{*}\approx10^{2}\lambda m$. Thus a typical adult skyrmion mass, 
for say $\lambda=10$, is 
$10^{3}$ times atomic mass.  
Finally, the equation of motion for $\lambda(t)$ reads
\begin{equation}
2\lambda\ddot\lambda+{\dot\lambda}^{2}+c=0
\label{lambdaeq},
\end{equation} 
where $c=(29-20S)\pi$. 
In this equation lengths are again scaled to the correlation length $\xi$ and time is 
scaled to the correlation time $\tau=2m\xi^2/\hbar$. 
In these units the condition of validity of this equation is $\lambda\gg1$. 
For the initial conditions $\lambda(0)=\lambda_{0}$ and ${\dot\lambda}(0)=0$, 
we find that 
\begin{equation}
\lambda=\lambda_{0}(1-{c\over 4\lambda_{0}^{2}}t^{2})
\label{lambdasol}
\end{equation} 
is a solution to Eq. (\ref{lambdaeq}). 
This formula shows that adult skyrmions shrink with a rate 
$\gamma_{\rm adult}\approx\sqrt{c}/2\lambda_{0}\tau$ which indicates that spin-1/2 
skyrmions 
shrink almost twice as fast as spin-1 skyrmions. 
In both cases one can make the shrinking rate smaller by exciting larger 
skyrmions. 
For realistic estimates of these shrinking rates, we restore the units in the above expression for 
$\gamma_{\rm adult}$. For $^{87}$Rb spin-1/2 
condensate of central density 
$10^{-11}$cm$^{-3}$, the rate then reads $\gamma_{\rm adult}\approx18.06\xi/\lambda\;$sec$^{-1}$.

\subsection{Baby Skyrmions}
\label{baby}
The cruicial difference between adult skyrmions and baby skyrmions, apart 
from the size difference, is that density fluctuations for baby skyrmions 
are much more important than for the adult skyrmions.
The density depletion produced by the spin gradients  
increases for smaller skyrmions and thus one can not consider the density 
to be essentially uniform anymore.
Therefore, the linear-response approach followed in the previous subsection 
does not apply. 
To take the density fluctuations properly into account, we need to use the full energy 
functional. 
As already mentioned, it leads to a size dependence of the the 
skyrmion energy such that the 
skyrmion tends to shrink to zero size. 
The shrinking process, however, is now fundamentally different and 
more complicated 
than 
that for the adult skyrmions. 
In this case, while the skyrmion is shrinking the density 
becomes 
increasingly depleted in a region that forms a closed shell around the 
center of the skyrmion. 
For sufficiently small $\lambda$ the depletion will be so large that the 
atoms 
inside the closed shell are essentially isolated from the atoms outside the shell. 
This will considerably slow down the collapse of the skyrmion since the atoms 
within the shell need to tunnel over a potential barrier to 
escape to the other side of the shell and to enable the skyrmion to shrink further. 
It turns out that the size of the skyrmion at this stage is of the order of 
the correlation length and the lifetime of the skyrmion due 
to the above tunneling process can be much 
larger than the lifetime of the condensate itself.

\subsubsection{Density Profile and Texture}
\label{densityprofile}
With our ansatz for $\omega(\rho)$ in Eq. (\ref{omegaansatz}) 
the problem is significantly simplified 
since to determine the density profile we only have 
to solve Eq. (\ref{gp}), which is basically a 
nonlinear Schr\"odinger equation with some external potential 
$|{\bf \mbox{\boldmath $\nabla$}_{\mbox{\boldmath $\rho$}} }
\zeta(\mbox{\boldmath $\rho$})|^{2}$. 
Substituting our ansatz for $\omega(\rho)$ in Eq. (\ref{gradzeta2}), the 
latter takes the form
\begin{equation}
|{\mbox{\boldmath $\nabla$}}_{\mbox{\boldmath $\rho$}} \zeta(\mbox{\boldmath $\rho$})|^{2}
=\left\{
\begin{array}{cc}
32({\rho\over\lambda^{2}})^2{\left[3+2(\rho/\lambda)^4
+3(\rho/\lambda)^8\right]\over\left[1+(\rho/\lambda)^4\right]^4},&S=1/2\\\\
4(17+3\cos{\theta}^{2})({\rho\over\lambda^{2}})^{2}
{1\over\left[1+(\rho/\lambda)^{4}\right]^{2}},&S=1
\end{array}
\right.
\label{gradzeta22}.
\end{equation} 
This is an off-centered potential barrier with a maximum height of 
$24.3/\lambda^2$ at 
$r\approx0.68\lambda$, for $S=1/2$ 
and a maximum height equal to $3\sqrt{3}(17+3\cos^{2}{\theta})/4\lambda^{2}$ 
at $r=\lambda/3^{1/4}$ for $S=1$.
Using this form for $|{\bf \mbox{\boldmath $\nabla$} }\zeta(\mbox{\boldmath $\rho$})|^{2}$ 
we solve Eq. (\ref{gp}) numerically for one particular value of $\lambda$. 
Then we use the resulting density distribution $f(\mbox{\boldmath $\rho$})$ to calculate 
the energy for that particular value of $\lambda$. 
Performing the same calculation for different values of $\lambda$ 
we finally obtain the energy of the skyrmion as a function of $\lambda$, 
from which we can judge the stability of the skyrmion.

For the spin-1/2 case the density profile for various values of 
$\lambda$ is shown in Fig. \ref{profilea}. 
In the case of $S=1$ the calculation is more complicated due to the 
angular dependence of $|{\mbox{\boldmath $\nabla$}}_{\mbox{\boldmath $\rho$}}\zeta(\mbox{\boldmath 
$\rho$})|^{2}$. 
This complication is handled by noticing that 
$|{\mbox{\boldmath $\nabla$}}_{\mbox{\boldmath $\rho$}}\zeta(\mbox{\boldmath $\rho$})|^{2}$ can be 
rewritten in terms of 
$Y_{00}(\theta,\phi)$ and $Y_{20}(\theta,\phi)$ only, where 
$Y_{lm}(\theta,\phi)$ are the usual spherical harmonics. 
We thus also expand Eq. (\ref{gp}) in the spherical harmonics 
up to $l=2$ using 
\begin{equation}
\sqrt{f(\mbox{\boldmath $\rho$})}
=y_{0}(\rho)Y_{00}(\theta,\phi)
+y_{2}(\rho)Y_{20}(\theta,\phi)
\label{n}.
\end{equation}    
Then two coupled equations can be obtained by taking the $l=0$ and $l=2$ components of Eq. 
(\ref{gp}). 
Specifically, we substitute  
Eq. (\ref{n}) into Eq. (\ref{gp}) and multiply by $Y_{00}(\theta,\phi)$  
and $Y_{20}(\theta, \phi)$, respectively, and then perform an angular integration. 
The resulting equations take the form
\begin{equation}
-\nabla^{2}_{\rho}y_{0}+{16\over\lambda^{2}}{(\rho/\lambda)^{2}
\over(1+(\rho/\lambda)^{4})^{2}}
\left(4y_{0}+{1\over\sqrt{5}}y_{2}\right)
+{1\over4\pi}y_{0}^{3}
+{3\over4\pi}y_{0}y_{2}^{2}
+{\sqrt{5}\over14\pi}y_{2}^{3}
=y_{0}
\label{radialeq}
\end{equation}  
and
\begin{equation}
-\nabla^{2}_{\rho}y_{2}+{6\over \rho^{2}}y_{2}+{16\over\lambda^{2}}{(\rho/\lambda)^{2}
\over(1+(\rho/\lambda)^{4})^{2}}
\left({1\over\sqrt{5}}y_{0}+{58\over7}y_{2}\right)
+{3\over4\pi}y_{2}y_{0}^{2}
+{3\sqrt{5}\over14\pi}y_{0}y_{2}^{2}
+{15\over28\pi}y_{2}^{3}
=y_{2}
\label{angulareq},
\end{equation} 
where 
$\nabla^{2}_{\rho}={1\over \rho^{2}}{\partial\over\partial \rho}
\left(\rho^{2}{\partial\over\partial \rho}\right)$. 
The result of the numerical solutions of these two equations is presented 
in Fig. \ref{profileb}. From this figure it is clear that the anisotropic part of 
$f(\mbox{\boldmath $\rho$})$, which is represented by $y_{2}(\rho)$, 
is considerably smaller than the isotropic part. 
This result will be employed in the following to simplify the energy calculation 
by neglecting the angular dependence of $f(\mbox{\boldmath $\rho$})$. 
Note that for $S=1/2$ this is not an approximation. 

Having specified $\omega(r)$ enables also a detailed visualization of 
the skyrmion texture by calculating the average spin projections, 
i.e., $<S_{x}>={\zeta^{\dagger}({\bf r})}{S}_{x}\zeta({\bf r})$, 
$<S_{y}>$, and $<S_{z}>$. 
In Fig.  \ref{texturefig} we plot these quantities as a function of $r$ for 
the spin-1 case using, again, $\lambda=\xi$. Similar plots for the spin-1/2 
case can be found in our previous work \cite{us1}.
The skyrmion can be best visualized from its $\langle S_z\rangle$({\bf r}) 
component, which is shown in Appendix \ref{a} to be equal to
\begin{equation}
\langle S_{z}\rangle
=\cos^2{\theta}+\cos{\omega}\sin^{2}{\theta}.
\end{equation}
This expression is used to produce the two upper false-color 
figures in Fig. \ref{texturefig}. 
These figures show clearly that the skyrmion corresponds 
to a torodial region. Along the inner radius of the torus, the direction of the 
average spin vectors is opposite to that in the ground state. 
An interesting property shows up when we calculate the associated superfluid 
velocity 
${\bf v}_{\rm s}({\bf r})=-i\hbar\zeta^{\dagger}({\bf r})\mbox{\boldmath 
$\nabla$}\zeta({\bf r})/m$. 
This is given by
\begin{eqnarray}
{\bf v}_{\rm s}
=\hbar\left[\cos{\theta}{d\omega\over dr}\;{\hat{\bf r}}
-{1\over r}\sin{\theta}\sin{\omega}\;{\hat{\mbox{\boldmath 
$\theta$}}}
-{2\over r}\sin{\theta}\sin^{2}{(\omega/2)}\;{\hat{\mbox{\boldmath 
$\phi$}}}\right].
\end{eqnarray}
From the above two expressions we find that the superfluid velocity is such 
that the atoms are simultaneously rotating around the inner radius of the torus and 
around the $z$-axis. This corresponds to a spiraling motion 
around the inner radius of the torus. The speed is maximum in the center of the torus. 
Furthermore, the maximum depletion of the density takes place at the 
maximum of $|{\bf \mbox{\boldmath $\nabla$}}\zeta({\bf r})|^{2}$, 
which turns out to be in the center of the torus, where 
the spin vector is also completely flipped, i.e., $\langle S_z\rangle=-1$. 
In the case of a spin-1/2 condensate, the spin texture consists of two torii. 
The velocity field in this case is such that 
atoms spiral around the center of one torus clock wise 
and along the center of the other torus counter clock wise, 
but for both torii the atoms rotate around the $z$-axis in the same direction. 
The spins are completely flipped along the inner radii of the two torii. The density is mostly depleted 
along a radius that is slightly smaller than the inner radius of the larger torus. 
The shift is only approximately 3 percent of the size of the skyrmion.

\subsubsection{Equilibrium State}
\label{equilibriumstability}
The outcome of the calculation in the previous subsection 
is the skyrmion density profile as a function of $\lambda$. 
Therefore, the energy of the skyrmion, determined from Eq. (\ref{functional}), 
is now also a function of $\lambda$. 
The skyrmion will be energetically stable if this energy 
$E[\lambda]$ has a local or 
global minimum and the equilibrium size of the skyrmion will be 
equal to the value of 
$\lambda$ at that minimum. 
It should be noted that what we call the energy of the skyrmion 
is actually the grand-canonical energy. This implies that the density profile
is solved by using a fixed chemical potential for all values of 
$\lambda$. 
Consequently, the number of atoms associated with the skyrmion excitation, 
which equals $|\int d{\bf r} (n({\bf r})-n_{\infty})|$, 
is also $\lambda$-dependent. 
(The subtracted term is to cancel the divergent part coming from the fact that 
the system is infinite. In a confined system there is no need 
to make this subtraction.) 
It may therefore not be immediately clear that minimizing the grand-canonical 
energy is appropriate. 
In principle, we must minimize the true energy of the skyrmion at a fixed number of 
atoms. We now explicitly show though, that these procedures are equivalent.
Let us take the number of atoms in the system equal to $N$. This means that 
in the absence 
of the skyrmion $\int d{\bf r}\;n_{\infty}=N$. Suppose we now put in a skyrmion 
of size $\lambda$ in the condensate without affecting the asymptotic density, i.e., 
$n({\bf r}\rightarrow\infty)=n_{\infty}$. 
Due to the depletion of the density near the center of the skyrmion, 
the total number of atoms associated with this 
density profile is slightly less than $N$ by an amount $\Delta 
N(\lambda)=\int 
d{\bf r}(n_{\infty}-n({\bf r},\lambda) )$. 
Physically this means that to produce this skyrmion we have to remove 
$\Delta N(\lambda)\ll N$ atoms to the edge of the spinor condensate. 
If we, however, want to produce a skyrmion with the same number of 
atoms in the condensate, we have to adjust the asymptotic density. 
This implies that a reasonable approximation to the density profile is 
actually $(N/(N-\Delta N))n({\bf r},\lambda)$ because then
\begin{equation}
{N\over N-\Delta N}\int d{\bf r}\;n({\bf r},\lambda)=N{N-\Delta N(\lambda)
\over N-\Delta N(\lambda)}= N.
\end{equation}   
Using this density profile to calculate the true energy of the skyrmion and 
expanding the various terms in it up to first order in $\Delta N(\lambda)/N$, 
we reproduce exactly the grand-canonical energy $E[\lambda]$. 
Thus, we conclude that 
minimizing the 
canonical energy with a fixed number of atoms is indeed equivalent to minimizing the 
grand-canonical 
energy at a fixed chemical potential and thus 
with a changing number of atoms. The second method is clearly more convenient 
numerically since 
it is difficult to keep the number of atoms fixed for each value of $\lambda$.

In Fig. \ref{energy} we show our final result of this calculation, 
where we have taken $n({\bf r})$ to be isotropic, since in the previous section we 
saw that the angular-dependence of $n({\bf r})$ can indeed be neglected. 
In detail, we have plotted here 
$E[n(r)]-\mu n(r)-(E[n_{\infty}]-\mu n_{\infty})$ 
against $\lambda$. Again, we subtract the divergent part due to 
the infiniteness of the system. 
Unfortunately, both of the two curves do not have a minimum 
for any finite $\lambda$ and only contain a global minimum at $\lambda=0$. 
This means that, according to this figure, 
the skyrmion is energetically unstable, i.e., if the skyrmion is created with a finite size 
it will ultimately shrink to zero size. 
This process is clearly seen in Fig. \ref{profilea} where we plot the skyrmion 
density profile 
for different values of $\lambda$. 
In the next subsection we examine this shrinking process 
for the baby skyrmions more closely. 
In particular, we again look at the time scale over which this shrinking 
occures and find that for current experimental situations it is of the 
order of, or even longer than, the lifetime of the condensate itself. 

Finally, we mention here some remarks about the effect of 
a homogeneous external magnetic field and the slight differences in the scattering lengths 
of the different hyperfine spin states. 
In the presence of a homogeneous magnetic field $\bf B$, 
it will be energetically unfavorable to flip spins. 
Therefore, this will lead to a further reduction in the spin of the skyrmion and,
thus, not to a stabilization.
A slight difference in the various scattering lengths $\Delta a$, as appropriate for the 
spin-1/2 $^{87}$Rb condensate \cite{JILA3}, leads to an additional term in the energy expression for 
the 
spinor condensate of the form $(4\pi\Delta a\hbar^2/m)\int d{\bf r}\;n({\bf r})
\langle S_z\rangle({\bf r})$, 
which has exactly the same effect as an external magnetic field. 
It can not stabilize the skyrmion either.

\subsubsection{Shrinking Rate}
\label{nonequilibriumstability}
The goal of this section is to estimate the time scale over which the skyrmion 
shrinks to zero size. 
A key element in understanding the dynamics of this process is the behavior 
of the potential barrier produced by the spin gradients 
$|{\bf \mbox{\boldmath $\nabla$}}\zeta({\bf r})|^{2}$ 
when $\lambda$ goes to zero. 
This potential is written down explicitly in Eq. (\ref{gradzeta22}). 
When $\lambda$ is decreasing the location of the 
off-centered peak of this potential is approaching the origin as $\lambda$ and 
its height is increasing as $1/\lambda^{2}$. 
This potential is thus an isotropic three-dimensional repulsive shell that is 
increasing in strength as it becomes smaller. 
We denote from now on the atoms inside this shell by the ``core'' atoms and 
the ones 
outside by the ``external'' atoms.
We have seen in the previous subsection that, in principle, the skyrmion is unstable and 
that it is energetically favorable for the skyrmion to shrink to zero size 
irrespective of the initial size. 
Let us therefore consider a skyrmion of a large initial size such that the height 
of the corresponding potential barrier is less than the chemical potential 
$\mu=n_{\infty}T^{2B}$ 
of the system. 
The density distribution will be almost uniform everywhere except in 
the region where the 
barrier has its maximum height. In this region the density profile will 
be depleted only slightly. 
When the skyrmion starts to shrink the depletion of the density becomes larger, since 
the 
height of the barrier increases. 
This is achieved physically by transporting the 
core atoms over the potential barrier to the external region. 
So the central density $n({\bf 0})$ decreases while the skyrmion is shrinking, as can 
be seen in Fig. \ref{profilea}.  
At a certain size the height of the barrier becomes equal to the 
chemical 
potential of the atoms at large distances. 
As the skyrmion shrinks even further the barrier height actually exceeds the chemical potential. 
At this stage the shrinking process slows down significantly since the core atoms 
will now have to tunnel through a potential barrier. 
The rate of this process will be characterized by the usual WKB tunneling rate, which 
we calculate now. 

To calculate this tunneling rate we assume that the skyrmion has shrunk to a size for which the barrier 
is so high that the overlap between the wave 
functions 
of the core atoms and the external atoms is exponentially small. 
The tunneling rate will then be determined by the chemical potential of the core 
atoms, as well as the barrier height. 
We consider the situation that there are $N_{\rm core}$ core atoms with a chemical 
potential $\mu_{\rm core}$ less than the height of the barrier. 
Note that physically we must always have that $\mu_{\rm core}\geq\mu$ since the collapse of the 
skyrmion will squeeze the atoms inside the core. For small $r/\lambda$ we notice from Eq. 
(\ref{gradzeta22}) that the 
potential $(\hbar^2/2m)|{\bf \mbox{\boldmath $\nabla$}}\zeta({\bf r})|^{2}$ can 
be approximated by a harmonic potential well with 
a characteristic frequency $\omega_{0}$ that equals 
$\sqrt{48}\hbar/m\lambda^{2}$ for $S=1/2$ and 
$\sqrt{32}\hbar/m\lambda^{2}$ for $S=1$. 
It thus has a characteristic length $l=\lambda/\sqrt{96}$ for $S=1/2$ 
and $l=\lambda/8$ for $S=1$. 
This allows us to use a Thomas-Fermi 
approximation, the validity of which will be discussed below, to calculate 
$\mu_{\rm core}$. 
Moreover, the tunneling rate is calculated using the following WKB expression \cite{Stoof}
\begin{equation}
\gamma_{baby}\approx{\omega_0\over 2\pi}
\exp{\left[-2\int_{r_1}^{r_2}dr
\sqrt{|{\bf \mbox{\boldmath $\nabla$}}\zeta|^{2}-
{2m\mu_{\rm core}/\hbar^2}}\right]}
\label{tunneling}.
\end{equation}                                                                 
The radial points $r_1$  
and $r_2$ are the points where $(\hbar^2/2m)|{\bf \mbox{\boldmath $\nabla$}}\zeta({\bf r})|^{2}$ and 
$\mu_{\rm core}$ 
intersect. 
The use of a Thomas-Fermi approximation is justified 
when the mean-field interaction energy of the core atoms is larger than the spacing 
between the lowest energy levels of the harmonic potential well. Specifically, 
the ratio $2Na/l=2\sqrt{96}Na/\lambda$ should be bigger than 1. 
Figure \ref{rate_baby}a shows the equilibrium size of the baby skyrmion, 
which is calculated by minimizing the total energy while keeping the number of the 
core atoms constant, as a function of the number of the core atoms $N_{\rm core}$. 
From this figure we observe that the above 
ratio equals approximately 1 for $N_{\rm core}=4$ and it increases for larger 
$N_{\rm core}$. 
For experimental situations, Fig.  \ref{rate_baby}b shows that the skyrmion 
can live for a time ranging from seconds to hundreds of seconds, which is 
long enough to be observed. It should be noted that the above expression, 
which is based on a variational approach, 
somewhat overestimates the lifetime of the skyrmion.  
However, this is not crucial for our purposes, since by shrinking slightly 
the lifetime of the skyrmion increases considerably.

\section{Skyrmion Dynamics}
\label{dynamicproperties}
A very important dynamical variable of the
skyrmion arises from the fact that the Euler-Lagrange equations for the skyrmion
spin texture is invariant under a space independent rotation of the average
local spin $\langle {\bf S} \rangle({\bf r})$ around the magnetic field
direction $\hat{\bf B}$ \cite{rene}. Mathematically, this means that if
the spinor $\zeta^{\rm sk}({\bf r})$ describes a skyrmion, then
$\exp\{-i\vartheta \hat{\bf B} \cdot {\bf S}\} \zeta^{\rm sk}({\bf r})$
describes also a skyrmion with the same winding number and energy.
The dynamics of the variable
$\vartheta(t)$ associated with this symmetry is determined by the full action
for the spin texture $S[\zeta] = \int dt (T[\zeta] - E[\zeta])$, 
where $E[\zeta]$ and $T[\zeta]$ are given by Eqs. (\ref{egrad}) and (\ref{t}) with $n_{\infty}$ 
replaced by the density profile $n({\bf r})$ of the skyrmion and $\chi({\bf r}-{\bf r}')$ should now be 
interpreted as the exact density-density correlation function in the presence of the skyrmion. 
Hence, substituting $\zeta({\bf r},t)
    = \exp\{-i\vartheta(t) \hat{\bf B} \cdot {\bf S}\} \zeta^{\rm sk}({\bf r})$
and making use of the conservation of total particle number $N$ to introduce
the change of the average local spin projection on the magnetic field
$\langle \Delta S_{||} \rangle({\bf r})
    = \hat{\bf B} \cdot \langle {\bf S} \rangle({\bf r}) - NS$
induced by the skyrmion,
we obtain, apart from an unimportant boundary term, that
the dynamics of the rotation angle $\vartheta(t)$ is determined by the action
\begin{equation}
\label{action}
S[\vartheta] = \int dt~
  \left\{ \frac{\partial \vartheta(t)}{\partial t}
             \hbar \langle \Delta S^{\rm tot}_{||} \rangle
    + \frac{1}{2} I \left( \frac{\partial \vartheta(t)}{\partial t} \right)^2
  \right\}~,
\end{equation}
where $\langle \Delta S^{\rm tot}_{||} \rangle$ is the change of the total
spin along the magnetic field direction and the ``moment of inertia'' of the
skyrmion equals
\begin{equation}
I = \hbar^2 \int d{\bf r} \int d{\bf r}'~
       \langle \Delta S_{||} \rangle({\bf r})
          \chi({\bf r},{\bf r}') \langle \Delta S_{||} \rangle({\bf r}')~.
\end{equation}
For an adult skyrmion this integral can be performed by substituting $\langle \Delta S_{||} 
\rangle=\langle 
S_z\rangle-NS$ and using Eq. (\ref{response}) for $\chi({\bf r},{\bf r}')$. The 
explicit 
expressions for $\langle S_z\rangle$ can be found in Appendix \ref{a}. An even simpler 
expression for 
$\chi({\bf r},{\bf r}')$ can be optained by neglecting the gradient terms in Eq. 
(\ref{response}). Then
\begin{equation}
\chi({\bf r},{\bf r}')={m\over4\pi\hbar^2a}\delta({\bf r}-{\bf r}')~.
\end{equation}  
Using this approximation we find,
\begin{equation}
I=\left\{
\begin{array}{lr}
{53\pi\over40\sqrt{2}}{\lambda^3\over a}m,&S=1/2\\\\
{14\pi\over15\sqrt{2}}{\lambda^3\over a}m,&S=1
\end{array}
\right.~.
\end{equation}

The importance of this result is twofold. First, from the action
in Eq. ~(\ref{action}) we see that at the quantum
level the hamiltonian for the dynamics of the wave function
$\Psi(\vartheta,t)$ becomes
\begin{equation}
H = \frac{1}{2I} \left( \frac{\hbar}{i}
                          \frac{\partial}{\partial \vartheta}
                        - \hbar \langle \Delta S^{\rm tot}_{||} \rangle
                 \right)^2~.
\end{equation}
Therefore, the ground state wave function is given by
$\Psi_0(\vartheta) = e^{iK\vartheta}/\sqrt{2\pi}$, with $K$ an
integer that is as close as possible to $\langle \Delta S^{\rm
tot}_{||} \rangle$. In this way we thus recover the fact that
according to quantum mechanics the total number of spin-flips
associated with the skyrmion texture must be an integer. More
precisely, we have actually shown that the many-body wave function
describing the skyrmion is an eigenstate of the operator $\hat{\bf
B} \cdot {\bf S}^{\rm tot}$ with eigenvalue $NS - K$. Note that
physically this is equivalent to the way in which ``diffusion'' of
the overall phase of a Bose-Einstein condensate leads to the
conservation of particle number \cite{li,henk2}.

Furthermore, the existence of this internal degree of freedom
becomes especially important when we deal with more than one skyrmion in
the condensate. In that case every skyrmion can have its own orientation
and we expect the interaction between two skyrmions to have a
Josephson-like contribution proportional to $\cos(\vartheta_1-\vartheta_2)$.
As a result the phase diagram of a gas of skyrmions can become extremely
rich and contain both quantum as well as classical, i.e., nonzero
temperature, phase transitions \cite{rene}. In this context, it is interesting
to mention two important differences with the situation in the quantum Hall
effect. First, the fact that the spin projection $K$ of the skyrmion is an
integer shows that these excitations have an integer spin and are therefore
bosons \cite{FR}. In the quantum Hall case the skyrmions are fermions with
half-integer spin, due to the presence of a topological term in the action
$S[\zeta]$ for the spin texture \cite{wilczek}.
Second, in a spinor Bose-Einstein condensate the skyrmions are not pinned by
disorder and are in principle free to move. Both differences
will clearly have important consequences for the many-body physics of a
skyrmion gas.

Focusing again on the single skyrmion dynamics, we can make the
last remark more quantitative by using the ansatz $\zeta({\bf
r},t)= \zeta^{\rm sk}({\bf r}-{\bf u}(t))$ for the texture of a
moving skyrmion, which is expected to be accurate for small
velocities $\partial {\bf u}(t)/\partial t$. Considering for illustrative purposes again only the 
isotropic approximation, we find
in a similar way as before that the action for the center-of-mass
motion of the skyrmion becomes
\begin{equation}
S[{\bf u}] = \int dt~
 \frac{1}{2} M \left( \frac{\partial {\bf u}(t)}{\partial t} \right)^2~,
\end{equation}
where the mass is
now simply given by
\begin{equation}
M = \frac{m^2}{3} \int d{\bf r} \int d{\bf r}'~ \chi({\bf r},{\bf r}')
       \langle {\bf v}_{\rm s} \rangle({\bf r}) \cdot
           \langle {\bf v}_{\rm s} \rangle({\bf r}')
\end{equation}
in terms of the superfluid velocity of the spinor condensate
${\bf v}_{\rm s}({\bf r})$. 
Note that in the anisotropic spin-1 case this mass is in principle a tensor, but with only very small 
nondiagonal components.

Similar to the moment of inertia, we can calculate this mass explicitly for adult skyrmions. The result 
is
\begin{equation}
M=\left\{
\begin{array}{lc}
{19\pi\over18\sqrt{2}}{\lambda\over a}m,&S=1/2\\\\
{2\sqrt{2}\pi\over3}{\lambda\over a}m,&S=1
\end{array}
\right.~.
\end{equation}
The skyrmions thus indeed behave in this respect as ordinary
particles. In contrast to Eq. ~(\ref{action}) there thus appears no
term linear in $\partial {\bf u}(t)/\partial t$ in the action
$S[{\bf u}]$. This is a result of the fact that we have performed
all our calculations at zero temperature. In the presence of a
normal component, we anticipate the appearance of such a linear
term with an imaginary coefficient. This will lead to damping of
the center-of-mass motion of the skyrmion. It is interesting to
note that if we perform the same analysis for a vortex in a scalar
condensate, we find, due to the singular nature of the superfluid
velocity field, even at zero temperature an additional term linear
in the velocity of the vortex. In a quasi two-dimensional
situation, it is in fact proportional to ${\bf u}(t) \times
\partial {\bf u}(t)/\partial t$. This precisely results in the
well-know Euler dynamics of vortices.

\section{Conclusion}
\label{conclusion}
One important conclusion of this work is that skyrmions in a ferromagnetic 
Bose-Einstein condensate are energetically unstable. 
However, we have also shown that the time scale on which the skyrmion shrinks can be of the order of, 
or even larger, than the life time of the condensate. 
It turns out that there are two very different mechanisms 
for the shrinking of the skyrmion. The first occurs for skyrmions with sizes much larger than the 
correlation length of the gas. In this case the shrinking rate is of the 
order of 
seconds for realistic experimental parameters. The second case concerns smaller 
skyrmions with sizes of the order or less than the correlation length of the 
system. 
In this case the shrinking rate is determined by the tunneling rate from the core 
of the 
skyrmion to the uniform part of the system. The tunneling takes place through a 
potential 
barrier developed by spin deformations. 
The typical time scale for such a shrinking process is of the order of 10 - 100 s. 

Although we have considered a homogeneous discussion, the inclusion of a trap in the 
system 
will not change the main conclusions of this work. 
The presence of a trap will have the effect that the uniform density 
$n_{\infty}$ 
will be now a position-dependent quantity. 
Therefore, for our calculations to be valid for a trapped gas, the size of the 
skyrmion 
must be much less than the typical length scale for the density gradients of the confined 
condensate. 
Thus the condition $\lambda\ll R$ must be met, 
where $R$ is the size of a spherical condensate. 
For typical experimental conditions we can use the Thomas-Fermi 
approximation to the size of the condensate\cite{chris}, which reads 
$R=\sqrt{2/\pi}(Na/l)^{1/5}l$ where $N$ is the number of atoms in the 
condensate 
and $l=\sqrt{\hbar/m\omega_{0}}$ 
is the characteristic length of a harmonic trap 
of frequency $\omega_{0}$. 
Scaling $R$ to the correlation length $\xi$, the former takes the form 
$R=4N^{1/5}(l/a)^{4/5}$. For a $^{87}$Rb condensate with $N=10^{6}$ atoms and 
a trap frequency 
of the order of $2\pi\times100$ rad/s this gives $R\approx500$. 
This gives an estimate for the maximum size of a skyrmion such that the above 
condition 
remains valid. For example, our theory for adult skyrmions is applicable for the 
size range of 
$\lambda=10\xi$ up to say $50\xi$. 

Another important remark about a confined gas is 
that skyrmions will tend to move towards the surface of the Bose-Einstein 
condensate since the density there is lower and thus 
their energy will be less. 
Apparently, this leads to a decrease in the shrinking rate since for 
both adult and baby skyrmions the shrinking time is proportional to $1/n_{\infty}$, see Eq. 
(\ref{lambdasol}) and the 
discussion afterwards, 
and Eq. (\ref{tunneling}) and the discussion before it.
However, our theory for the  shrinking rates of skyrmions clearly breaks 
down on the surface since there the typical length scale for the density gradients due to the trap are 
of the same order as the correlation length.  
 
Although skyrmion-antiskyrmion pairs can be created by the Kibble mechanism 
in a temperature quench or by sufficiently shaking up the condensate, a more 
controlled way of creating a skyrmion can be achieved by using a magnetic 
field configuration in which the fictitious magnetic field is always pointing 
radially outward and its magnitude increases monotonically from zero at the 
origin to a maximum value for large distances from the origin  \cite{anglin}. 
Applying this 
field configuration for such a long time that the spins at large distances 
have precessed exactly twice around the local magnetic field, creates a 
single skyrmion. Of course, for a real magnetic field the above configuration 
requires the use of magnetic monopoles, but for a fictitious magnetic field 
it can be achieved by appropriately tailoring the detuning, 
the polarization 
and the intensity of two pulsed Raman lasers. 
The required spatial dependence of the detuning can be created experimentally 
by separating the centers of the magnetic traps for the two spin species 
along the $z$-axis \cite{JILA2}. 
Furthermore, the desired behavior of the Rabi frequency can be achieved 
by making with the first Raman laser two standing waves in the $x$ and $y$ 
directions that are both polarized perpendicular to the $z$-axis. 
For the other Raman laser we only need to require that it produces a 
traveling wave with a polarization that has a nonzero projection on the $z$-axis, 
since we want to realize a $\Delta m=0$ transition in this case. 
In the above geometry the skyrmion is created exactly in the plane where the 
detuning vanishes and in the nodes of the first Raman laser. 
Note that since the distance between these nodes is generally much bigger 
than the correlation length, we create in this manner a large skyrmion that will 
start to shrink but ultimately selfstabilizes at a smaller size. 
Once created, the skyrmion can be 
easily observed by the usual expansion experiments that have recently also been 
used to observe vortices \cite{dalibard}. Similar as with vortex rings, we then observe an 
almost complete depletion of the condensate in a ring around the position of 
the skyrmion.

\section*{Acknowledgements}
We would like to thank Eric Cornell and Jan Smit for useful and stimulating 
discussions. 
This work is supported by the Stichting voor Fundamenteel Onderzoek der Materie 
(FOM), 
which is financially supported by the 
Nederlandse Organisatie voor Wetenschappelijk Onderzoek (NWO).

\newpage

\section*{Figure Captions}

\begin{figure}[p]
\begin{center}
\end{center}
\caption{
Schematic figure representing the action of the spin rotation operator for a maximally symmetric 
skyrmion. 
At a position $\bf r$ the average spin vector is pointing initially in the 
$z$-direction. The spin rotation operator rotates the spin vector around $\bf r$ 
by an angle $\omega(r)$.   
}
\label{schematic}
\end{figure}


\begin{figure}[p]
\begin{center}
\end{center}
\caption{
The density profile for a skyrmion in a spin-1/2 condensate for different values of 
$\lambda$. These curves are the numerical solutions of Eq. (\ref{gp}) 
using Eq. (\ref{gradzeta22}). 
}
\label{profilea}
\end{figure}

\begin{figure}[p]
\begin{center}
\end{center}
\caption{
Density profile for a skyrmion in a spin-1 spinor condensate with a size of $\lambda=\xi$. 
The solid curve represents the isotropic part of $n(\bf r)$ and the dashed 
curve 
represents the anisotropic part. These are the numerical solution of 
Eqs. (\ref{radialeq}) and (\ref{angulareq}). 
}
\label{profileb}
\end{figure}

\begin{figure}[p]
\begin{center}
\end{center}
\caption{
False-color figures representing the average spin $\langle {\bf S}\rangle({\bf r})$ for a skyrmion in a 
spin-1 condensate in 
different Cartesian planes. The size of the skyrmion is taken here to be equal to $\lambda=\xi$. The 
$\langle S_y\rangle$ figures can be obtained 
from the $\langle S_x\rangle$ figures by using the axial symmetry of the texture.  
}
\label{texturefig}
\end{figure}

\begin{figure}[p]
\begin{center}
\end{center}
\caption{
Energy of a skyrmion as a function of its size. 
Plotted with the solid (dashed) curve is the energy for a spin-1 (spin-1/2) 
spinor condensate. The dotted asymptotes are the large $\lambda$ limit 
of the energy, which is linear in $\lambda$ and 
is given by the spin gradient term in the energy 
functional only. 
Energy is in units of the correlation energy 
$E_{\xi}=(4\pi\hbar^2/2m\xi^2)$. 
}
\label{energy}
\end{figure}

\begin{figure}[p]
\begin{center}
\end{center}
\caption{
Shown are (a) the baby skyrmions size as a function of the number of core atoms and 
(b) the shrinking rate of skyrmion as a function of the number of core atoms. 
This calculation was performed for a $^{87}$Rb spin-1/2 condensate with a 
scattering length of $a=5.4$ nm.  
}
\label{rate_baby}
\end{figure}

\section*{Figures}

\appendix

\section{Details of the 
calculation of \lowercase{$\langle \uppercase{S}_{x}\rangle$, $\langle \uppercase{S}_{y}\rangle$, 
$\langle 
\uppercase{S}_{z}\rangle$, 
$\zeta^{\dagger}({\bf r})
\mbox{\boldmath $\nabla$}\zeta({\bf r})$}, and \lowercase{$|\mbox{\boldmath 
$\nabla$}\zeta|^{2}({\bf 
r})$} }
\label{a}
We start by expanding the exponential operator in Eq. (\ref{spinor}), taking into 
account the simplification expressed in Eq. (\ref{simplicity}). 
Inserting the explicit forms of the spin-1/2 and spin-1 matrices given by 
\cite{book}
\begin{equation}
{{\bf S}}=\left\{ 
\begin{array}{lr}
{1\over 2}
\left(\begin{array}{cc}
0&1\\
1&0
\end{array}\right)
{\hat{\bf x}}
+
{i\over2}
\left(\begin{array}{cc}
0&-1\\
1&0
\end{array}\right)
{\hat{\bf y}}
+
{1\over2}
\left(\begin{array}{cc}
1&0\\
0&-1
\end{array}\right)
{\hat{\bf z}}
,& S={1\over2}\\
{1\over\sqrt{2}}
\left(\begin{array}{ccc}
0&1&0\\
1&0&1\\
0&1&0
\end{array}\right)
{\hat{\bf x}}
+
{i\over\sqrt{2}}
\left(\begin{array}{ccc}
0&-1&0\\
1&0&-1\\
0&1&0
\end{array}\right)
{\hat{\bf y}}
+
\left(\begin{array}{ccc}
1&0&0\\
0&0&0\\
0&0&-1
\end{array}\right)
{\hat{\bf z}},
& S=1
\end{array}
\right.
\end{equation}
and noting that 
\begin{math}
({\hat{\bf S}}\cdot{\hat{\bf r}})^{k}
\end{math} 
equals ${{\bf S}}\cdot{{\bf r}}$ for $k=$odd, and equals 
$({{\bf S}}\cdot{\hat{\bf r}})^{2}$ for $k=$even enables a 
resuming of the odd 
and even powers of the exponent separately. As a result, one can put this 
exponential operator 
in the following simple form
\begin{eqnarray}
\exp{\left\{-{i\over S}{\mbox{\boldmath $\Omega$}}({\bf r})\cdot 
{{\bf S}}\right\}}
&=&\exp{\left\{-{i\over S}\omega(r){\hat{\bf r}}\cdot{{\bf 
S}}\right\}}\nonumber\\
&=&\left\{
\begin{array}{lr}
{\bf 1}\cos{(\omega(r))}-i({\hat{\bf r}}\cdot{{\bf 
S}})\sin{(\omega(r))},&S={1\over2}\\\\
{\bf 1}-({\hat{\bf r}}\cdot{{\bf S}})^{2}\left[
1-\cos{(\omega(r))}\right]-i({\hat{\bf 
r}}\cdot{{\bf S}})
\sin{(\omega(r))},&S=1
\end{array}
\right.
\label{operator},
\end{eqnarray}
where $\bf 1$ is the identity matrix.
It is then straightforward to derive the expressions for 
the important physical quantities such as the average spin components 
$\langle S_{x}\rangle=\zeta^{\dagger}({\bf r}){S}_{x}\zeta({\bf r})$, 
$\langle S_{y}\rangle$ and $\langle S_{z}\rangle$, and the 
superfluid velocity 
${\bf v}_{\rm s}=-i\hbar\zeta^{\dagger}({\bf r})\mbox{\boldmath 
$\nabla$}\zeta({\bf r})/m$.  
For $S=1/2$ and 1, these quantities take the form
\begin{equation}
\langle S_{x}\rangle/S
=\sin{\theta}\sin{\phi}\sin{(\omega/S)}
+\sin{(2\theta)}\cos{(\phi)}\sin^2{(\omega/2S)}~,
\end{equation}
\begin{equation}
\langle S_{y}\rangle/S
=-\sin{\theta}\cos{\phi}\sin{(\omega/S)}
+\sin{(2\theta)}\sin{(\phi)}\sin^2{(\omega/2S)}~,
\end{equation}
\begin{equation}
\langle S_{z}\rangle/S
=\cos^{2}(\theta)+\cos{(\omega/S)}\sin^{2}{\theta}~,
\end{equation}
and
\begin{eqnarray}
{\bf v}_{\rm s}
=\hbar\left[\cos{\theta}{d\omega\over dr}\;{\hat{\bf r}}
-{S\over r}\sin{\theta}\sin{(\omega/S)}\;{\hat{\mbox{\boldmath 
$\theta$}}}
\nonumber
-{2S\over r}\sin{\theta}\sin^{2}{(\omega/2S)}\;{\hat{\mbox{\boldmath $\phi$}}}\right]~. 
\end{eqnarray}
We can now derive the expression for the spin gradient energy  
$\hbar^{2}|\mbox{\boldmath $\nabla$}\zeta({\bf r})|^{2}/2m$ as Eq. 
(\ref{gradzeta22})
shows apart from the constant $\hbar^{2}/2m$.




\begin{references}
\bibitem{scat} In principle, this is only true if the inter and intra
               scattering lengths of the two condensate components are equal,
               which is a good approximation for $^{87}$Rb
               \cite{JILA2}.
\end{references}
\end{document}